\documentstyle[11pt,epsfig,rotate]{article}
\input{epsfig}
\begin{document}
\def\Tr{\,{\rm Tr}\,}
\def\beq{\begin{equation}}
\def\eeq{\end{equation}}
\def\beqa{\begin{eqnarray}}
\def\eeqa{\end{eqnarray}}
\begin{titlepage}
\vspace*{-1cm}
\noindent
\phantom{bla}
\\
\vskip 2.0cm
\begin{center}
{\Large {\bf Electromagnetic Corrections to $K \to \pi\pi$ I --
Chiral Perturbation Theory}}
\end{center}
\vskip 1.5cm
\begin{center}
{\large Vincenzo Cirigliano$^a$, John F. Donoghue$^b$
and Eugene Golowich$^b$} \\
\vskip .15cm
$^a$ Dipartimento di Fisica dell'Universit\`a and I.N.F.N. \\
Via Buonarroti,2 56100 Pisa (Italy) \\
vincenzo@het.phast.umass.edu \\

\vskip .15cm

$^b$ Department of Physics and Astronomy \\
University of Massachusetts \\
Amherst MA 01003 USA\\
donoghue@het.phast.umass.edu \\
gene@het.phast.umass.edu \\

\vskip .3cm
\end{center}
\vskip 1.5cm
\begin{abstract}
\noindent
An analysis of electromagnetic corrections to the
(dominant) octet $K \to \pi \pi$ hamiltonian
using chiral perturbation theory is carried out.  
Relative shifts in amplitudes at the several per 
cent level are found.  
\end{abstract}
\vfill
\end{titlepage}

\section{Introduction}
In this paper, we present a formal analysis of
electromagnetic (EM) radiative corrections to
$K \to \pi \pi$ transitions.\footnote{We restrict
our attention to EM corrections only and omit
consideration of $m_u \ne m_d$.  See however Ref.~\cite{cw}}
Only EM corrections
to the dominant octet nonleptonic hamiltonian are
considered.  Such corrections modify not only
the original $\Delta I = 1/2$ amplitude but also induce
$\Delta I = 3/2, 5/2$ contributions as well.
By the standards of particle physics, this subject is
very old~\cite{old}.  Yet, there exists in the literature no
satisfactory theoretical treatment.  This is due largely
to complications of the strong interactions at low energy.
Fortunately, the modern machinary of the Standard Model,
especially the method of chiral lagrangians, provides the means
to perform an analysis which is both correct and
structurally complete.  That doing so requires no fewer than
{\it eight} distinct chiral langrangians is an indication
of the complexity of the undertaking.

There is, however, a problem with the usual chiral
lagrangian methodology.  The cost of implementing its calculational
scheme is the introduction of many unknown constants,
the finite counterterms associated with the regularization of
divergent contributions.  As regards EM corrections to
nonleptonic kaon decay, it is impractical to presume that
these many unknowns will be inferred phenomenologically in
the reasonably near future, or perhaps ever.  As a consequence,
in order to obtain an acceptable phenomenological description,
it will be necessary to proceed beyond the confines
of strict chiral perturbation theory.  In a previous
publication~\cite{cdg}, we succeeded in accomplishing
this task in a limited context, $K^+ \to \pi^+ \pi^0$
decay in the chiral limit.  We shall extend this work to a
full phenomenological treatment of the $K \to \pi \pi$ decays
in the next paper~\cite{cdg2} of this series.

The proper formal analysis, which is the subject of this paper,
begins in Sect.~2 where we briefly describe the construction
of $K \to \pi\pi$ decay amplitudes in the presence of
electromagnetic corrections.  In Section 3, we begin to
implement the chiral program by
specifying the collection of strong and electroweak
chiral lagrangians which bear on our analysis.  The
calculation of $K \to \pi\pi$ decay amplitudes is
covered in Section 4 and our concluding remarks appear
in Section 5.

\begin{figure}
\vskip .1cm
\hskip 0.2cm
\epsfig{figure=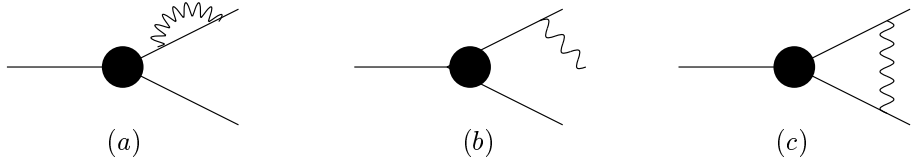,height=0.8in}
\caption{Some electromagnetic contributions.\hfill
\label{fig:f1}}
\end{figure}

\section{Electromagnetism and the $K \to \pi\pi$ Amplitudes}
There are three physical $K \to \pi\pi$ decay
amplitudes,\footnote{The invariant amplitude ${\cal A}$ is defined
via $_{\rm out}\langle \pi\pi|K\rangle_{\rm in}
= i (2\pi)^4 \delta^{(4)}(p_{\rm out} - p_{\rm in}) \left(
i {\cal A}\right)$.}
\beq
{\cal A}_{K^0 \to \pi^+ \pi^-} \equiv {\cal A}_{+-}\ ,
\qquad
{\cal A}_{K^0 \to \pi^0 \pi^0} \equiv {\cal A}_{00}\ , \qquad
{\cal A}_{K^+ \to \pi^+ \pi^0} \equiv {\cal A}_{+0}\ \ .
\label{a0}
\eeq
We consider first these amplitudes in
the limit of exact isospin symmetry and then identify
which modifications must occur in the presence
of electromagnetism.

In the $I = 0,2$ two-pion isospin basis, it
follows from the unitarity constraint that
\beqa
{\cal A}_{+-} \ &=&\  A_0 e^{i \delta_0} + \sqrt{1\over 2} A_2 e^{i
\delta_2} \ \ ,
\nonumber \\
{\cal A}_{00} \ &=& \ A_0 e^{i \delta_0} - \sqrt{2} A_2
e^{i \delta_2} \ \ ,
\label{a1} \\
{\cal A}_{+0} \ &=& \ {3 \over 2} A_2 e^{i \delta_2} \ \ .
\nonumber
\eeqa
The phases $\delta_0$ and $\delta_2$ are just the $I = 0,2$ 
pion-pion scattering phase shifts (Watson's theorem), and 
in a CP-invariant world the moduli $A_0$ and $A_2$ are 
real-valued.  The large ratio $A_0/A_2 \simeq 22$ is 
associated with the $\Delta I = 1/2$ rule.

\begin{figure}
\vskip .1cm
\hskip 2.8cm
\epsfig{figure=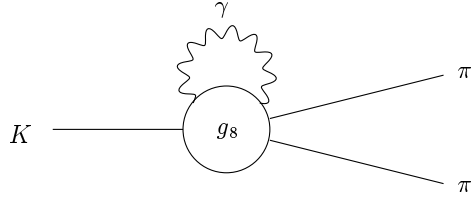,height=1.0in}
\caption{Leading electromagnetic correction to $K \to \pi \pi$.\hfill
\label{fig:f2}}
\end{figure}

When electromagnetism is turned on, several new features appear:
\begin{enumerate}
\item Charged external legs experience mass
shifts ({\it cf} Fig.~\ref{fig:f1}(a)).
\item Photon emission ({\it cf}
Fig.~\ref{fig:f1}(b)) occurs off charged external legs.  This 
effect is crucial to the cancelation of infrared singularities.
\item Final state coulomb rescattering ({\it cf}
Fig.~\ref{fig:f1}(c)) occurs in $K^0 \to \pi^+ \pi^-$.
\item There are structure-dependent hadronic effects,
hidden in Fig.~1 within the large dark vertices.  In this paper,
we consider the leading contributions (see Fig.~\ref{fig:f2})
which arise from corrections to the $\Delta I = 1/2$ hamiltonian.
\item There will be modifications of the isospin symmetric
unitarity relations and thus extensions of Watson's theorem.
\end{enumerate}
Any successful explanation of EM corrections to $K \to \pi\pi$ decays
must account for all these items.

An analysis~\cite{cdg4} of the unitarity constraint which
allows for the presence of electromagnetism yields
\beqa
{\cal A}_{+-} &=& \left( A_0 + \delta A_0^{\rm em} \right)
e^{i(\delta_0 + \gamma_0)} + { 1 \over \sqrt{2}} \left( A_2 +
\delta A_2^{\rm em}  \right)
e^{i(\delta_2 + \gamma_2)}
\ , \nonumber \\
{\cal A}_{00} &=& \left( A_0 + \delta A_0^{\rm em}  \right)
e^{i(\delta_0 + \gamma_0)} - \sqrt{2} \left( A_2 +
\delta A_2^{\rm em}  \right)
e^{i(\delta_2 + \gamma_2)} \ \ ,
\label{a6} \\
{\cal A}_{+0} &=& {3 \over 2} \left( A_2
+ \delta A_2^{+{\rm em}}  \right)
e^{i (\delta_2 + \gamma_2')} \ \ ,
\nonumber
\eeqa
to be compared with the isospin invariant expressions
in Eq.~(\ref{a1}).  This parameterization holds for 
the IR-finite amplitudes, whose proper definition is discussed 
later in Sect.~4.3.  Observe that the shifts
$\delta A_2^{+{\rm em}}$ and $\gamma_2'$ in ${\cal A}_{+0}$
are distinct from the corresponding shifts in ${\cal A}_{+-}$
and ${\cal A}_{00}$.  This is a consequence of
a $\Delta I = 5/2$ component induced by electromagnetism.
In particular, the $\Delta I = 5/2$ signal can be recovered via
\beq
{\cal A}_{5/2} = { \sqrt{2} \over 5} \left[
{\cal A}_{+-} - {\cal A}_{00} - \sqrt{2} {\cal A}_{+0} \right] \ \ .
\label{a6f}
\eeq

\section{Chiral Lagrangians}
The preceding section has dealt with aspects of the
$K \to \pi\pi$ decays which are free of hadronic
complexities.  In this section and the next, we use
chiral methods to address these structure-dependent
contributions.  The implementation of chiral symmetry
via the use of chiral lagrangians provides a logically
consistent framework for carrying out a perturbative analysis.

In chiral perturbation theory, the perturbative quantities of
smallness are the momentum scale $p^2$ and the mass scale
$\chi = 2 B_0 {\bf m}$, where ${\bf m}$ is the quark mass
matrix. In addition, we work to first order in
the electromagnetic fine structure constant $\alpha$,
\beq
{\cal A}_i = {\cal A}_i^{(0)} + \alpha {\cal A}_i^{(1)} + \dots \ \ .
\label{c1}
\eeq
Our goal is to determine the
${\cal O}(\alpha)$ components $\alpha {\cal A}_i^{(1)}$.
The fine structure constant thus represents a second
perturbative parameter, and we consider contributions
of chiral orders ${\cal O}(e^2 p^0)$ and ${\cal O}(e^2 p^2)$,
\beq
\alpha {\cal A}_i^{(1)} \ \equiv \ {\cal A}_i^{(e^2 p^0)} +
{\cal A}_i^{(e^2 p^2)} \ \ .
\label{c1a}
\eeq

We shall restrict our attention to just the leading
electromagnetic corrections to the $K \to \pi\pi$
amplitudes.  Since the weak $\Delta I= 1/2$ amplitude
is very much larger than the $\Delta I= 3/2$ amplitude,
our approach is to consider only electromagnetic corrections
to $\Delta I= 1/2$ amplitudes.  As a class these arise via
processes contained in Fig.~\ref{fig:f2}, where $g_8$ is
the octet weak coupling defined below in Eq.~(\ref{c6}).

We adopt standard usage in our chiral analysis, taking
the matrix $U$ of light pseudoscalar fields and its
covariant derivative $D_\mu U$ as
\beq
U \equiv \exp(i \lambda_k \Phi_k /F_\pi) \ \ (k = 1,\dots,8) \ ,
\qquad
D_\mu U \equiv \partial_\mu U + i e [Q,U] A_\mu \ \ ,
\label{c2}
\eeq
where $Q = ~diag~(2/3,-1/3,-1/3)$ is the quark charge
matrix and $A_\mu$ is the
photon field.  The remainder of this section summarizes
the eight distinct effective lagrangians (strong,
electromagnetic, weak and electroweak) needed in the analysis.

\subsection{Strong and Electromagnetic Lagrangians}
In the $\Delta S = 0$ sector, we shall employ the
strong/electromagnetic lagrangian
\beq
{\cal L}^{(2)}_{\rm str} = {F^2_0\over 4}
\Tr \left( D_\mu U D^\mu U^\dagger \right) +
{F^2_0\over 4} \Tr \left(\chi U^\dagger + U\chi ^\dagger\right) \ ,
\label{c3}
\eeq
where $F$ is the pseudoscalar meson decay constant in
lowest order.  ${\cal L}^{(2)}_{\rm str}$ will be used to
produce ${\cal O}(e^0 p^2)$ and ${\cal O}(e^1 p^1)$ vertices
in our calculation.

The lagrangian ${\cal L}^{(2)}_{\rm str}$ will generate
(via tadpole diagrams) strong self-energy effects on the
external legs in the $K \to \pi \pi$ transitions.  In order
to regularize these divergent contributions, one employs
the lagrangian~\cite{gl2} ${\cal L}^{(4)}_{\rm str}$.  It
is not necessary to write out this well-known set of operators,
but simply to point out that the resulting wave function
renormalization factors $Z_\pi$ and $Z_K$ obey
\beq
{1 \over F_\pi^2 F_K} \ = \ {Z_\pi \sqrt{Z_K} \over F^3} \ \ ,
\label{c3a}
\eeq
up to logarithms.  This explains the presence of $F_\pi^2 F_K$
in formulae such as Eqs.~(\ref{d7}),(\ref{d8}) in Section 4.

Two other nonweak effective lagrangians enter the calculation.
The first is associated with electromagnetic effects at chiral
order ${\cal O}(e^2 p^0)$,
\beq
{\cal L}_{\rm ems}^{(0)} = g_{\rm ems}
\Tr \left( Q U Q U^\dagger \right) \ \ ,
\label{c4}
\eeq
where the coupling $g_{\rm ems}$ is fixed (in lowest chiral 
order) from the pion electromagnetic mass splitting,
\beq
g_{\rm ems}  = {F_\pi^2 \over 2}~ \delta M_\pi^2 \ \ .
\label{c5}
\eeq
The second extends the description to chiral order
${\cal O}(e^2 p^2)$.  We need only the following subset 
of the lagrangian given in Ref.~\cite{ur}, 
\beqa
& & {\cal L}_{\rm ems}^{(2)} = F^2 e^2 \bigg[ \kappa_1
\Tr  \left( D_\mu U D^\mu U^\dagger \right) \cdot \Tr Q^2
\nonumber \\
& & \phantom{xxxxxxxx} ~ + \kappa_2 \Tr \left( D_\mu U D^\mu U^\dagger \right) \cdot
\Tr \left( Q U Q U^\dagger \right)
\nonumber \\
& & \phantom{xxxxxxxx}~ + \kappa_3 \left( \Tr \left( D_\mu U^\dagger  Q U \right)
\cdot \Tr \left( D^\mu U^\dagger  Q U \right) \right.
\nonumber \\
& & \left. \phantom{xxxxxxxxxxxxx} +
\Tr \left( D_\mu U Q U^\dagger \right)
\cdot \Tr \left( D^\mu U Q U^\dagger \right) \right)
\nonumber \\
& & \phantom{xxxxxxxx}~ + \kappa_4 \Tr \left( D_\mu U^\dagger  Q U \right)
\cdot \Tr \left( D^\mu U Q U^\dagger \right)
\label{c5a} \\
& & \phantom{xxxxxxxx}~ + \kappa_5 \left( \Tr \left( D_\mu U^\dagger
D_\mu U  Q \right) + \Tr \left( D_\mu U D^\mu U^\dagger Q \right) \right)
\nonumber \\
& & \phantom{xxxxxxxx}~ + \kappa_6 \Tr \left( D_\mu U^\dagger D^\mu U
Q U^\dagger Q U + D_\mu U D^\mu U^\dagger Q U Q U^\dagger \right)
\bigg] \ \ .
\nonumber
\eeqa
Although the finite parts of the coefficients $\kappa_1, \dots,
\kappa_6$ remain unconstrained, see however
Refs.~\cite{bp,bu,mo} for model determinations.

\subsection{Weak Lagrangians}
The $|\Delta S| = 1$ octet lagrangian begins at chiral order $p^2$,
\beq
{\cal L}_8^{(2)} \ = \ g_8 \Tr \left( \lambda_6 D_\mu U D^\mu U^\dagger
\right) \ \ ,
\label{c6}
\eeq
with $g_8 \simeq 6.7 \cdot 10^{-8}~F_\pi^2$ fit~\cite{kmw} 
from $K \to \pi \pi$ decay rates.  We use this to generate
${\cal O}(e^0 p^2)$, ${\cal O}(e^1 p^1)$ and ${\cal O}(e^2 p^0)$
vertices.

Two chiral lagrangians will serve to provide counterterms for
removing divergent contributions.  The first~\cite{ekw} is the octet
$|\Delta S| = 1$ lagrangian at chiral order $p^4$,
\beqa
{\cal L}_8^{(4)} &=& N_5 \Tr \lambda_6 \bigg[ \left( U \chi^\dagger 
+ \chi U^\dagger  \right) \partial_\mu U \partial^\mu U^\dagger
+ \partial_\mu U \partial^\mu U^\dagger \left( U \chi^\dagger 
+ \chi U^\dagger  \right) \bigg]
\nonumber \\
&+& N_6 \Tr \lambda_6 U \partial_\mu U^\dagger  \cdot \Tr \left(
\chi^\dagger \partial^\mu U - \chi \partial^\mu U^\dagger  \right)
\nonumber \\
&+& N_7 \Tr \lambda_6 \left( U \chi^\dagger + \chi U^\dagger \right)
\cdot \Tr \partial_\mu U \partial^\mu U^\dagger
\nonumber \\
&+& N_8 \Tr \lambda_6 \partial_\mu U \partial^\mu U^\dagger \cdot
\Tr \left( U^\dagger \chi + \chi^\dagger U \right)
\nonumber \\
&+& N_9 \Tr \lambda_6 \bigg[ \partial_\mu U \partial^\mu U^\dagger 
\left( \chi U^\dagger - U \chi^\dagger  \right) - 
\left( \chi U^\dagger - U \chi^\dagger  \right) \partial_\mu U 
\partial^\mu U^\dagger \bigg]
\nonumber \\
&+& N_{10} \Tr \lambda_6 \left( U \chi^\dagger U \chi^\dagger 
+ \chi U^\dagger \chi U^\dagger + U \chi^\dagger \chi U^\dagger 
\right)
\nonumber \\
&+& N_{11} \Tr \lambda_6 \left( U \chi^\dagger + \chi U^\dagger 
\right) \cdot \Tr \left( U^\dagger \chi + \chi^\dagger U \right)
\nonumber \\
&+& N_{12} \Tr \lambda_6 \left( U \chi^\dagger U \chi^\dagger 
+ \chi U^\dagger \chi U^\dagger - U \chi^\dagger \chi U^\dagger 
\right)
\nonumber \\
&+& N_{13} \Tr \lambda_6 \left( U \chi^\dagger - \chi U^\dagger 
\right) \cdot \Tr \left( U \chi^\dagger - \chi U^\dagger \right) \ \ .
\label{c7}
\eeqa
At present, little is known of the finite parts of the 
couplings $\{ N_k \}$.

\subsection{Electroweak Lagrangians}
The $|\Delta S| = 1$ lagrangian at chiral order ${\cal O}(e^2 p^0)$ is
\beq
{\cal L}_{\rm emw}^{(0)} = g_{\rm emw}
\Tr \left( \lambda_6 U Q U^\dagger \right) \ \ ,
\label{c8}
\eeq
where $g_{\rm emw}$ is an {\it a priori} unknown coupling
constant.  It has been calculated recently in Ref.~\cite{cdg},
\beq
g_{\rm emw} = \left( -0.62 \pm 0.19 \right) g_8 \delta M_\pi^2 \ \ .
\label{c9}
\eeq
We note in passing that despite the presence of just one
charge matrix $Q$ the lagrangian of Eq.~(\ref{c8}) indeed
describes ${\cal O}(e^2)$ effects.  A second factor of $Q$
could be decomposed into a combination of the unit matrix and
the $3 \times 3$ matrix ${\hat Q}= ~diag~(1,0,0)$.
The contribution from ${\hat Q}$ would vanish, leaving the
form of Eq.~(\ref{c8}).

The second operator that we use to provide counterterm
contributions is the $|\Delta S| = 1$ lagrangian at chiral order
${\cal O}(e^2 p^2)$.  In terms of the notation
$L_\mu \equiv i U \partial_\mu U^\dagger$, we have
\beqa
& & {\cal L}_{emw}^{(2)} = e^2 g_8 \bigg[
s_1 \Tr \lambda_6 [ Q, L_\mu Q L^\mu ]_+
\nonumber \\
& & + s_2 \Tr \lambda_6 \left( Q U Q U^\dagger L_\mu L^\mu
+ L_\mu L^\mu U Q U^\dagger Q \right)
\nonumber \\
& & + s_3 \Tr \lambda_6 [ Q, L_\mu U Q U^\dagger L^\mu ]_+
+ s_4 \Tr \lambda_6 [ L_\mu , U Q U^\dagger]_+  \cdot
\Tr U Q U^\dagger L^\mu
\nonumber \\
& & + s_5 \Tr \lambda_6 \left( Q U Q U^\dagger \chi U^\dagger
+ U \chi^\dagger U Q U^\dagger Q\right)
\nonumber \\
& & + s_6 \Tr \lambda_6 [ \chi , U^\dagger ]_+
\cdot \Tr U Q U^\dagger Q
\label{c11}  \\
& & + s_7 \Tr \lambda_6 \left( U Q U^\dagger Q \chi U^\dagger
+ U \chi^\dagger Q U Q U^\dagger \right)
\nonumber \\
& & + s_8 \Tr \left( \lambda_6 \partial_\mu U
\partial^\mu U^\dagger \right) \cdot \Tr Q^2
\nonumber \\
& & + s_9 \Tr \left( \lambda_6 \partial_\mu U
\partial^\mu U^\dagger \right) \cdot \Tr U Q U^\dagger Q
\bigg] \ \ . \nonumber
\eeqa
The first six operators in the above list appear in
Ref.~\cite{der}.  The remaining three are also required
for our analysis.  To our knowledge, none of
the divergent or finite parts of the $\{ s_n \}$ are
yet known.  

\section{Calculation of Leading EM Corrections}
\begin{figure}
\vskip .1cm
\hskip 0.2cm
\epsfig{figure=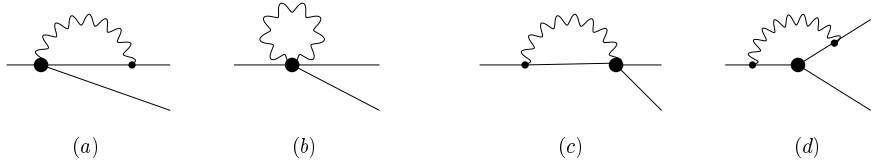,height=0.8in}
\caption{Explicit photon contributions in $K^+ \to \pi^+ \pi^0$.\hfill
\label{fig:f3}}
\end{figure}

The leading EM corrections arise from the processes of
Fig.~\ref{fig:f1} and Fig.~\ref{fig:f2}.
Contributions to Fig.~\ref{fig:f2}
occur in two distinct classes, those explicitly containing virtual
photons (Fig.~\ref{fig:f3}) and those with no explicit virtual
photons (Fig.~\ref{fig:f4}).  The latter are induced by
EM mass corrections and by insertions of $g_{\rm emw}$.  
In Figs.~\ref{fig:f3},\ref{fig:f4}, the larger bold-face 
vertices are where the weak interaction occurs.

The integrals which occur in our chiral analysis
are standard and already appear in the literature
({\it e.g.} see Ref.~\cite{gl1} or Ref.~\cite{gk1}).
It suffices here to point out that all divergent parts
of the one-loop integrals are ultimately expressible
in terms of the $d$-dimensional integral
\beq
A ( M^2 ) \equiv \int d{\tilde k}~
 ~ { 1 \over k^2 - M^2 }
= \mu^{d-4} \left[ -2i M^2 {\overline \lambda}
- { iM^2 \over 16 \pi^2} \log \left( {M^2 \over
\mu^2} \right) + \dots \right] \ \ ,
\label{bpp1}
\eeq
where $d{\tilde k} \equiv d^d k/(2\pi)^d$ is the integration measure,
$\mu$ is the scale associated with dimensional
regularization and ${\overline \lambda}$ is the singular quantity
\beq
{\overline \lambda} \equiv
{ 1 \over 16 \pi^2} \left[ {1 \over d - 4} - {1\over 2}
\left( \log {4\pi} - \gamma + 1 \right) \right] \ \ .
\label{bpp2}
\eeq
Each amplitude in the discussion to follow will be expressed as a
sum of a finite contribution and a singular term containing
${\overline \lambda}$.

\subsection{Summary of ${\cal O}(e^2)$ Amplitudes}
We begin with the ${\cal O}(e^2 p^0)$ amplitudes,
\beq
{\cal A}_{+-}^{(e^2 p^0)} = -{\sqrt{2} \over F_K F_\pi^2} \left(
g_8 \delta M_\pi^2 + g_{\rm emw} \right)~,
\quad {\cal A}_{00}^{(e^2 p^0)}  = 0 ~,
\quad {\cal A}_{+0}^{(e^2 p^0)}
= {{\cal A}_{+-}^{(e^2 p^0)}  \over \sqrt{2}} ~.
\label{d1}
\eeq
Although these have already been determined in Ref.~\cite{cdg},
we include them here for the sake of completeness.  They are
finite-valued and require no regularization procedure.

\begin{figure}
\vskip .1cm
\hskip 0.5cm
\epsfig{figure=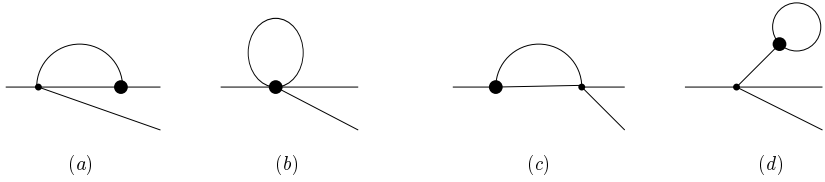,height=0.9in}
\caption{Diagrams without explicit photon contributions.\hfill
\label{fig:f4}}
\end{figure}

Next come the amplitudes of order $e^2 p^2$, expressed as
\beq
{\cal A}_i^{(e^2 p^2)}  = {\cal A}_i^{\rm (expl)} +
{\cal A}_i^{\rm (impl)} + {\cal A}_i^{\rm (ct)} \ \ .
\label{d2}
\eeq
The superscript `expl'
refers to Figs.~\ref{fig:f1}(a),(c) and Fig.~\ref{fig:f3}
where virtual photons are {\it explicitly} present, whereas
superscript `impl' refers to Fig.~\ref{fig:f4}
where EM effects are {\it implicitly} present via 
EM mass splittings and $g_{\rm emw}$ insertions.  The final 
term ${\cal A}^{\rm (ct)}$ is the counterterm amplitude.

\subsubsection{Diagrams with Explicit Photons}
We turn first to the class ${\cal A}^{\rm (expl)}$ of explicit
photonic diagrams.  For these contributions, it is consistent
to take meson masses in the isospin limit.  We find
\beqa
& & { F_K F_\pi^2 \over \sqrt{2} g_8} {\cal A}_{+-}^{\rm (expl)}
= \left( M_K^2 - M_\pi^2 \right) \cdot
\alpha {B}_{+-} (m_\gamma)
\nonumber \\
& & \phantom{xxxxx} + {\alpha \over 4 \pi}
\left[ 7 M_\pi^2 - 3 M_K^2 \left( \ln {M_\pi^2 \over \mu^2}
+ 1 \right) \right] - 6 \mu^{d-4} e^2 M_K^2 {\overline \lambda} \ \ ,
\nonumber \\
& & { F_K F_\pi^2 \over \sqrt{2} g_8 }
{\cal A}_{00}^{\rm (expl)} = 0 \ \ ,
\label{d7} \\
& & { F_K F_\pi^2 \over g_8 } {\cal A}_{+0}^{\rm (expl)}
= {\alpha \over 4 \pi} M_\pi^2
\left[ 7 - 3 \left( \ln {M_\pi^2 \over \mu^2}
+ 1 \right) \right] - 6 \mu^{d-4} e^2 M_\pi^2 {\overline \lambda} \ \ .
\nonumber
\eeqa

The quantity ${B}_{+-}$, which appears in the above
expression for ${\cal A}_{+-}^{\rm (expl)}$, is associated
with the processes of Figs.~\ref{fig:f1}(a),(c).  Due to
such processes, the weak decay amplitudes ${\cal A}_i$ will develop
infrared (IR) singularities in the presence of
electromagnetism. To tame such behavior, an IR
regulator is introduced and appears as a parameter in the
amplitudes.  For our work, this takes the form of a photon
squared-mass $m_\gamma^2$. ${B}_{+-}$ is given by
\beqa
{B}_{+-} ( m_\gamma^2) &=& {1 \over 4 \pi} \bigg[
2 a (\beta) \ln {M_\pi^2 \over m_\gamma^2} +
{1 + \beta^2 \over 2 \beta} h ( \beta)
+ 2 + \beta \ln {1 + \beta \over 1 - \beta}
\nonumber \\
& & \phantom{xxxxx} + i \pi \left( {1 + \beta^2 \over \beta}
\ln { M_K^2 \beta^2 \over m_\gamma^2} - \beta \right) \bigg] \ \ ,
\label{a50}
\eeqa
where
\beq
\beta = (1 - 4 M_\pi^2 /M_K^2 ) ^{1/2}
\label{a50a}
\eeq
and
\beqa
a (\beta) &=& 1 + {1 + \beta^2 \over 2 \beta}
\ln {1 - \beta \over 1 + \beta} \ \ ,
\nonumber \\
h ( \beta) &=& \pi^2 + \ln {1 + \beta \over 1 - \beta}
\ln {1 - \beta^2  \over 4 \beta^2 } + 2 f \left(
{1 + \beta \over 2 \beta} \right) - 2 f
\left( { \beta - 1 \over 2 \beta} \right) \ \ ,
\label{a51} \\
f(x) &=& - \int_0^x dt ~ {1\over t} \ln|1 - t| \ \ .
\nonumber
\eeqa
Notice that the function ${B}_{+-}$ is complex, and both its
real and imaginary parts have a logarithmic singularity as
$m_{\gamma} \rightarrow 0$. The solution to this problem is well
known; in order to get an infrared-finite decay rate, one has to
consider the process with emission of soft {\em real} photons,
whose singularity will cancel the one coming from soft {\em virtual}
photons.  We shall be more explicit on this point in Sect.~4.3.

The amplitudes
${\cal A}_{+-}^{\rm (expl)}$ and
${\cal A}_{+0}^{\rm (expl)}$ each contain an
additive divergent term (proportional to
${\overline \lambda}$) and also depend on the
arbitrary scale $\mu$ introduced in dimensional
regularization of loop integrals.  Both these features
will require the introduction of counterterms.

\subsubsection{Diagrams without Explcit Photons}

Next comes the class ${\cal A}^{\rm (impl)}$ of diagrams
in Fig.~\ref{fig:f4} not containing explicit photons.
For such contributions, one must be sure to include all
possible effects of chiral order ${\cal O}(e^2 p^0)$ and
${\cal O}(e^2 p^2)$ and treat the various terms in a consistent manner.
Thus for the contributions to Fig.~\ref{fig:f4},
isospin-invariant meson masses are used in amplitudes
involving ${\cal L}_{\rm emw}^{(0)} \times {\cal L}_{\rm str}^{(2)}$
and ${\cal L}_{\rm ems}^{(0)} \times {\cal L}_8^{(2)}$,
whereas electromagnetic mass splittings appear
in amplitudes involving
${\cal L}_{\rm str}^{(2)} \times {\cal L}_8^{(2)}$.
We write the results as sums of complex-valued finite
amplitudes ${\cal F}_i (\mu)$ and
divergent parts, essentially the amplitudes ${\cal D}_i$,
\beq
{\cal A}_i^{\rm (impl)} = {\cal R}e~{\cal F}_i (\mu) +
i {\cal I}m~{\cal F}_i (\mu) +
\mu^{d-4} {\cal D}_i ~{\overline \lambda} \ , \qquad
(i = +-, 00, +0) \ \ .
\label{d8}
\eeq
The scale-dependence in ${\cal F}_i (\mu)$ comes entirely
from its real part ${\cal R}e~{\cal F}_i (\mu)$.

We express the ${\cal R}e~{\cal F}_i$
in terms of dimensionless amplitudes $a_i^{\rm (impl)}$,
\beq
{\cal R}e~{\cal F}_i (\mu) = \eta_i {g_8 M_K^2 \over
F_\pi^2 F_K} a_i^{\rm (impl)}(\mu) \ \ ,
\label{d8p}
\eeq
with $\eta_{+-} = \eta_{00} = \sqrt{2}$, $\eta_{+0} = 1$.
Since the $a_i^{\rm (impl)}(\mu)$ coefficients have rather
cumbersome analytic forms, we find it most convenient to express
them in the compact form
\beq
a_i^{\rm (impl)}(\mu) = b_i^{\rm (M)} {\delta M_\pi^2 \over F^2} +
b_i^{\rm (g)} {g \over F^2} +
\bigg[ c_i^{\rm (M)} {\delta M_\pi^2 \over F^2} +
c_i^{\rm (g)} {g \over F^2} \bigg] \ln {\mu \over 1~{\rm GeV}}
\ \ ,
\label{d8q}
\eeq
where
\beq
g \ \equiv \ g_{\rm emw} / g_8 \ \ .
\label{d8r}
\eeq
The coefficients appearing in Eq.~(\ref{d8q}) are given in Table 1.

\begin{center}
\begin{tabular}{l|cccc}
\multicolumn{5}{c}{Table~1: {Values of Coefficients in
Eq.~(\ref{d8r})}}
\\ \hline\hline
 & $b_i^{\rm (M)}$ & $b_i^{\rm (g)}$ & $c_i^{\rm (M)}$
& $c_i^{\rm (g)}$ \\ \hline
$i = +-$ & $0.0160$  & $-0.0409$ & $-0.0078$ & $-0.0445$ \\
$i = 00$ & $-0.0170$ & $-0.0224$ & $-0.0371$ & $-0.0176$ \\
$i = +0$ & $-0.0265$ & $-0.0220$ & $-0.0419$ & $-0.0357$ \\
\hline
\end{tabular}
\end{center}

The finite functions also have imaginary parts
${\cal I}m~{\cal F}_i$ which arise entirely from the processes
in Fig.~\ref{fig:f4}(c).  From direct calculation we find
\beqa
& & { F_K F_\pi^2 F^2 \over \sqrt{2} g_8} {\cal I}m~{\cal F}_{+-}
= - {\beta \over 16 \pi} \left[ { M_K^2 \over 2}
\left( \delta M_\pi^2 + g \right) \right.
\nonumber \\
& & \left. \phantom{xxxxxxxxxxxxxxxxxxxxx}
+ \left( {1 \over \beta^2} - 2
\right) \left( M_K^2 - M_\pi^2 \right) \delta M_\pi^2 \right] \ ,
\nonumber \\
& & { F_K F_\pi^2 F^2 \over \sqrt{2} g_8 }
{\cal I}m~{\cal F}_{00} = - {\beta \over 16 \pi}
\left( M_K^2 -  M_\pi^2 \right)
\left[ \delta M_\pi^2 + g  \right.
\nonumber \\
& & \left. \phantom{xxxxxxxxxxxxxxxxxxxxx}
+ 2 { M_K^2 -  M_\pi^2 \over \beta^2}
{\delta M_\pi^2 \over M_K^2} \right] \ ,
\label{d8s} \\
& & { F_K F_\pi^2 F^2 \over g_8 } {\cal I}m~{\cal F}_{+0}
= {\beta \over 32 \pi} \left( M_K^2 - 2 M_\pi^2 \right)
\left( \delta M_\pi^2 + g \right) \ \ ,
\nonumber
\eeqa
where $\beta$ is defined in Eq.~(\ref{a50a}).  As a check
on our calculation, we have verified that the above results
are identical to those obtained from unitarity.

The singular parts of ${\cal A}_i^{\rm (impl)}$ are embodied
by the ${\cal D}$-functions,
\beqa
& & { F^2 F_K F_\pi^2 \over \sqrt{2} g_8 }
{\cal D}_{+-} = M_K^2 \left[ {1 \over 2}  \delta M_\pi^2 +
{13 \over 2}  g \right] +
M_\pi^2 \left[ 10 \delta M_\pi^2 + 7 g \right] \ \ ,
\nonumber \\
& & { F^2 F_K F_\pi^2 \over \sqrt{2} g_8 }
{\cal D}_{00} = \left( M_K^2 - M_\pi^2 \right) \left[
{19 \over 3}  \delta M_\pi^2 + 3 g \right] \ \ ,
\label{d8a} \\
& & { F^2 F_K F_\pi^2 \over g_8 } {\cal D}_{+0} =
M_K^2 \left[ {19 \over 3}  \delta M_\pi^2 +
{89 \over 18}  g \right] +
M_\pi^2 \left[ 4 \delta M_\pi^2 + {86 \over 9}  g \right] \ \ .
\nonumber
\eeqa
To arrive at the above, we
have used both the correspondence between $\delta M_\pi^2$ and
$g_{\rm ems}$ given in Eq.~(\ref{c5}) and also the relation
\beq
M_{\pi^\pm}^2 - M_{\pi^0}^2 \ = \ M_{K^+}^2 - M_{K^0}^2 \ \ ,
\label{d8b}
\eeq
in the evaluation of loop integrals.  The latter follows from
Dashen's theorem~\cite{rd} and is justified
since terms violating Dashen's theorem would begin to
contribute at the higher chiral order $e^2 p^4$.

\begin{figure}
\vskip .1cm
\hskip 0.2cm
\epsfig{figure=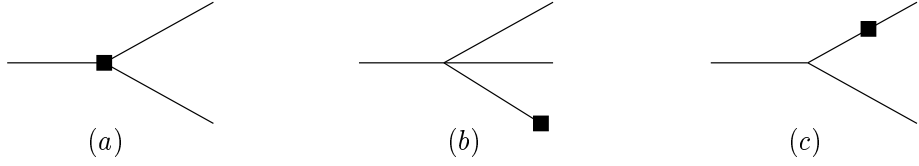,height=0.8in}
\caption{Counterterm contributions.\hfill
\label{fig:f5}}
\end{figure}

\subsection{The Regularization Procedure}
In order to cancel the singular ${\overline \lambda}$-dependence
in the $K \to \pi \pi$ amplitudes,
it is necessary to calculate all possible counterterm
amplitudes which can contribute. These enter in a variety of
ways, as shown in Fig.~\ref{fig:f5} where the small bold-face
square denotes the counterterm vertex.
For Figs.~\ref{fig:f5}(a),(b) the counterterm vertex
has $|\Delta S| = 1$ whereas in Fig.~\ref{fig:f5}(c) it
has $\Delta S = 0$.

\subsubsection{Counterterm Amplitudes}
Using the lagrangians ${\cal L}_8^{(4)}$, ${\cal L}_{\rm emw}^{(2)}$ and
${\cal L}_{\rm ems}^{(2)}$ we determine the counterterm amplitudes
to be
\beqa
& & { F^2 F_K F_\pi^2 \over \sqrt{2} g_8 }
{\cal A}_{+-}^{\rm (ct)} =
\nonumber \\
& & \phantom{xxxxx} M_K^2 \left( e^2 F^2 (X_1 - 4 U_1 - {8\over 3} U_2)
+ \delta M_\pi^2 ( 8 N_7 - 4 N_8 - 4 N_9) \right)
\nonumber \\
& & \phantom{xxxxx} + M_\pi^2 \left( e^2 F^2 (X_2 + 4 U_1 + {8\over 3} U_2)
- \delta M_\pi^2 ( 4 N_5 +  8 N_7 + 2 N_8) \right) \ \ ,
\nonumber \\
& & { F^2 F_K F_\pi^2 \over \sqrt{2} g_8 }
{\cal A}_{00}^{\rm (ct)} = \left( M_K^2 - M_\pi^2 \right) e^2 F^2
\bigg[ X_{00} - 4 U_1 - {8 \over 3} U_2 -
2 U_3 \bigg] \ , \label{d5} \\
& & { F^2 F_K F_\pi^2 \over g_8 } {\cal A}_{+0}^{\rm (ct)} =
M_K^2 \left( e^2 F^2 X_3 - \delta M_\pi^2 ( 4 N_5 + 4 N_8 ) \right)
\nonumber \\
& & \phantom{xxxxx} + M_\pi^2 \left( e^2 F^2 X_4
- \delta M_\pi^2 ( 2 N_8 + 4 N_9) \right) \ \ ,
\nonumber
\eeqa
where the $\{ N_i \}$ are coefficients in the $|\Delta S| =1$
lagrangian ${\cal L}_8^{(4)}$ of
Eq.~(\ref{c7}), the $\{ U_i \}$ are combinations of
coefficients in the $\Delta S =0$ lagrangian ${\cal L}_{\rm ems}^{(2)}$
of Eq.~(\ref{c5a}),
\beq
U_1 = \kappa_1 + \kappa_2 \ , \quad
U_2 = \kappa_5 + \kappa_6 \ , \quad
U_3 = -2 \kappa_3 + \kappa_4 \ ,
\label{d6}
\eeq
and the $\{ X_i \}$ are combinations of coefficients in the
$|\Delta S| =1$ lagrangian ${\cal L}_{\rm emw}^{(2)}$ of Eq.~(\ref{c5a}),
\beqa
X_1 &=& - {4 \over 9} s_1 - {1 \over 9} s_2 + {2 \over 9} s_3
+ {2 \over 3} s_5 - 4 s_6  + {2 \over 3} s_7 + s_8 + s_9  \ \ ,
\nonumber \\
X_2 &=&  {4 \over 9} s_1 - {2 \over 9} s_2 + {4 \over 9} s_3
+ {4 \over 3} s_5 + 4 s_6  - {2 \over 3} s_7 - s_8 - s_9  \ \ ,
\nonumber \\
X_3 &=& - {2 \over 3} s_1 - {1 \over 3} s_2 + {4 \over 3} s_4 +
{2 \over 3} s_5 + {2 \over 3} s_7 \ \ ,
\label{d6a} \\
X_4 &=&  {2 \over 3} s_1 + {2 \over 3} s_3 - {4 \over 3} s_4 +
{4 \over 3} s_5 - {2 \over 3} s_7 \ \ ,
\nonumber \\
X_{00} &=& {2 \over 9} \left( s_1 + s_2 + s_3 \right)
+ {2 \over 3} s_4 + s_8 + s_9 \ \ ,
\nonumber
\eeqa

\subsubsection{Removal of Divergences}
The counterterms themselves have finite and singular parts,
\beqa
N_i &=& n_i \mu^{d-4} {\overline \lambda} + N_i^{(r)} (\mu) \ \ ,
\nonumber \\
U_i &=& u_i \mu^{d-4} {\overline \lambda} + U_i^{(r)} (\mu) \ \ ,
\label{d6b} \\
X_i &=& x_i \mu^{d-4} {\overline \lambda} + X_i^{(r)} (\mu) \ \ .
\nonumber
\eeqa
The coefficients $n_i, u_i$ of the divergent parts of
$N_i,U_i$ have already been specified in the
literature~\cite{ekw,ur} and hence the $\mu$-dependences
of $N_i^{(r)}$, $U_i^{(r)}$ are known from the
renormalization group equations.  We infer the $x_i$ coefficients
in this paper by canceling divergences in the
${\cal O}(e^2p^2)$ amplitudes.  Upon combining results 
obtained thus far, we find the new results
\beqa
x_{00} &=& - {1\over 3} {\delta M_\pi^2 \over e^2 F^2}
- 3 { g \over e^2 F^2} \ \ ,
\nonumber \\
x_1 &=& 3 + {27\over 2} {\delta M_\pi^2 \over e^2 F^2}
- {13 \over 2} { g \over e^2 F^2} \ \ ,
\nonumber \\
x_2 &=& 3 - 18 {\delta M_\pi^2 \over e^2 F^2}
- 7 { g \over e^2 F^2} \ \ ,
\label{d9c} \\
x_3 &=& - {7\over 3} {\delta M_\pi^2 \over e^2 F^2}
- {89 \over 18}  { g \over e^2 F^2} \ \ ,
\nonumber \\
x_4 &=& 6 - 2 {\delta M_\pi^2 \over e^2 F^2}
- {86 \over 9}  { g \over e^2 F^2} \ \ ,
\nonumber
\eeqa
where we recall $g \equiv g_{\rm emw}/g_8$.

\subsection{Removal of Infrared Singularities}
Removal of the infrared divergence from the
expression for the decay rate is achieved by taking into account
the process ${K}^0 \to \pi^+ \pi^- (n \gamma)$.  For soft
photons, whose energy is below the detector resolution $\omega$, this
process cannot be experimentally distinguished from ${K}^0 \to
\pi^+ \pi^-$, so the observable quantity involves the inclusive sum over
the $\pi^+\pi^-$ and $\pi^+\pi^- (n \gamma)$ final states.

At the order we are working, it is sufficient to consider just the
emission of a single photon. The amplitude for the radiative decay is
given in lowest order by
\begin{equation}
\left. {\cal A}_{+ - \, \gamma} = e \frac{\sqrt{2} g_{8}}{F_K F^{2}_{\pi}}
\, (M_{K}^{2} - M_{\pi}^{2}) \, \left(
\frac{ \epsilon \cdot  p_{+}}{ q \cdot p_{+}}
 - \frac{ \epsilon \cdot  p_{-}}{ q \cdot p_{-}} \right)  \ \ ,
\right.
\label{v1}
\end{equation}
where $\epsilon$ and $q$ are the polarization and momentum of the
emitted photon. \\ The infrared-finite observable decay rate is
\begin{equation}
\left. \Gamma_{+ -} ( \omega ) =  \Gamma_{+  -}  \  +  \
\Gamma_{+  - \, \gamma} (\omega) \ \ \ \ ,
  \right.
\label{v2}
\end{equation}
where
\begin{eqnarray}
\Gamma_{+ -} & = & \frac{1}{2 M_{K}} \, \int \, d \Phi_{+ -} \
 | {\cal A}_{+ -} |^{2} \ \ , \\
\Gamma_{+ \, - \, \gamma} (\omega) & = & \frac{1}{2 M_{K}} \,
\int_{E_{\gamma} < \omega} \, d \Phi_{+ \, - \, \gamma}  \
 | {\cal A}_{+ \, - \, \gamma} |^{2} \ \ ,
\end{eqnarray}
and $ d \Phi_{k}$ is the differential phase space factor for each
process. The infrared divergent (IRD) part of $\Gamma_{+ -}$ is seen to be
\begin{equation}
\left. \Gamma_{+ -}^{\rm (IRD)} =  \frac{1}{2 M_{K}} \,
\left[ \frac{\sqrt{2} g_{8}}{F_K F^{2}_{\pi}}
\, (M_{K}^{2} - M_{\pi}^{2}) \right]^{2}
\int \, d \Phi_{+ -} \  \  2 \alpha \, {\cal R}e
 {B}_{+ -} (m_{\gamma})   \right. \ \ .
\label{v3}
\end{equation}
Equation~(\ref{v3}) displays explicitly the singularity and shows
that the imaginary part of ${B}_{+ -} (m_{\gamma})$ has no observable
effect at this order. This result has been shown to be true to all
orders in $\alpha$~\cite{sw1,yfs}.  For $\Gamma_{+ \, - \, \gamma}
(\omega)$ we get the following expression, up to terms of order
$\omega /M_{K}$,
\beq
\Gamma_{+ \, - \, \gamma} (\omega) =  \frac{1}{2 M_{K}} \,
\left[ \frac{\sqrt{2} g_{8}}{F_K F^{2}_{\pi}}
\, (M_{K}^{2} - M_{\pi}^{2}) \right]^{2}
\int \, d \Phi_{+ - } \, \ {I}_{+-}(m_{\gamma},\omega) \ \ ,
\label{x1}
\eeq
where
\beq
{I}_{+-}( m_\gamma , \omega ) = {\alpha \over \pi} \left[
a (\beta) \ln \left( { m_\gamma \over 2 \omega } \right)^2 +
F ( \beta ) \right] \ \ ,
\label{x2}
\eeq
with
\beqa
& & F ( \beta ) = {1 \over \beta} \ln {1 + \beta \over 1 - \beta}
+ {1 + \beta^2 \over 2 \beta} \bigg[
2 f ( -\beta) - 2 f (\beta) + f \left({1 + \beta \over 2}\right)
\nonumber \\
& & \phantom{xxxxx} - f \left({1 - \beta \over 2}\right) + {1 \over 2}
\ln {1 + \beta \over 1 - \beta} ~\ln (1 - \beta^2 ) +
\ln 2 ~\ln {1 - \beta \over 1 + \beta} \bigg] \ \ .
\label{x3}
\eeqa
From these explicit expressions of ${B}_{+  - } (\omega)$ and
${I}_{+-} (m_{\gamma},\omega)$ it is easy to see
that the combination $2 \alpha\, {\cal R}e  {B}_{+ - }
(m_{\gamma})  + {I}_{+-}(m_{\gamma},\omega)$
does not depend on the infrared regulator $m_{\gamma}$. However, this
combination has a dependence on the experimental resolution $\omega$.
To obtain a meaningful prediction therefore requires 
knowledge of the experimental treatment of soft photons.  A careful 
discussion of this point will appear in Ref.~\cite{cdg4}.  

A generalization of the above considerations beyond the order ${\cal
O}(e^2 p^2)$ in ChPT leads to the following parameterization,
\beq
\Gamma_{+ -} (\omega) =  \frac{1}{2 M_{K}} \,
\int  d \Phi_{+ -} \ G_{+-}(\omega) ~\big| {\cal A}_{+ -}^{(0)} \,  + \,
\alpha \, {\cal A}_{+ -}^{(1)} \big|^{2} \ \ ,
\label{v4}
\eeq
where to first order in $\alpha$, 
\beq
G_{+-}(\omega) = 
1 + 2 \alpha \, {\cal R}e {B}_{+ -} (m_{\gamma}) \,
+ \, {I}_{+-} \, (m_{\gamma},\omega) \ \ .
\label{v4a}
\eeq
With the prescription of dropping the term proportional to 
${B}_{+ -}$ in the photonic loop contribution, the 
electromagnetic amplitude $ \alpha {\cal A}_{+ -}^{(1)} $
can be read from Eqs.~(\ref{d1}),(\ref{d7}),(\ref{d8}),(\ref{d5}).

\subsection{The Finite Amplitudes}
The physical amplitudes will be complex-valued functions,
as dictated by unitarity.  The real parts are obtained by combining
the finite loop amplitudes (Eq.~(\ref{d7}) for ${\cal A}_i^{\rm (expl)}$
and Eqs.~(\ref{d8p}),(\ref{d8q}) along with Table 1
for ${\cal A}_i^{\rm (impl)}$) with
the counterterm amplitudes of Eq.~(\ref{d5}),
\beq
{\cal R}e~{\cal A}_i^{(e^2 p^2)} \ = \ \eta_i {g_8 M_K^2 \over
F_\pi^2 F_K} \left[ {\cal R}e~a_i^{\rm (loop)} + a_i^{\rm (ct)} \right] \ \ .
\label{d15}
\eeq
In order to make the scale-dependence of
${\cal R}e~a_i^{\rm (loop)}$ explicit, we write
\beq
{\cal R}e~a_i^{\rm (loop)} \ = \ b_i + c_i \ln {\mu \over 1~{\rm GeV}}
\ \ .
\label{d16}
\eeq
Numerical determination of the above quantities will depend on 
$g_8$ (obtained from Ref.~\cite{kmw}), 
$\delta M_\pi^2$ and $g_{\rm emw}$ (given in Eq.~(\ref{c9})).
We obtain the central values
\beqa
\begin{array}{l}
b_{+-} = 11.8 \cdot 10^{-3} \ , \\
b_{00} =  - 0.5 \cdot 10^{-3} \ , \\
b_{+0} = - 1.3 \cdot 10^{-3} \ ,
\end{array}
\qquad
\begin{array}{l}
c_{+-} = 7.1 \cdot 10^{-3} \ ,\\
c_{00} = -3.9 \cdot 10^{-3} \ , \\
c_{+0} = -2.7 \cdot 10^{-3} \ \ .
\end{array}
\label{d17}
\eeqa

The imaginary parts of the physical amplitudes can be
either determined from unitarity or
read off from Eqs.~(\ref{d8}),(\ref{d8s}).  Of
most interest is the EM shift in ${\cal A}_2$,
as only it receives the ${\cal A}_0 /{\cal A}_2$
($\Delta I = 1/2$) enhancement,
\beqa
\delta ({\cal I}m~{\cal A}_2^{\rm em}) &=& { \beta \over 32 \pi} \left[
{\cal A}_2^{(e^2 p^0)} {\cal T}_2^{(e^0 p^2)} +
{\cal A}_0^{(e^0 p^2)} {\cal T}_{02}^{(e^2 p^0)} \right.
\nonumber \\
& & \left. \phantom{xxxxxxx}
- {2 \sqrt{2} \over 3 \beta^2} ~ {\delta M_\pi^2 \over M_K^2}
{\cal A}_0^{(e^0 p^2)} {\cal T}_2^{(e^0 p^2)} \right] \ ,
\label{d17a}
\eeqa
where ${\cal T}_2^{(e^0 p^2)}$ and ${\cal T}_{02}^{(e^2 p^0)}$ 
are pion-pion T-matrix elements in the isospin basis. 
The above three contributions have physically distinct origins;
the first involves the direct effect of electromagnetism on the $I=2$
decay amplitude, the second arises from final state scattering in
which electromagnetism induces leakage from $I = 0$ to $I = 2$,
and the third is due to the shift in two-pion phase space
produced by the electromagnetic mass shift~\cite{cdg4}.

\section{Final Results and Concluding Remarks}
Despite the presence of many unknown finite counterterms,
it is possible to apply the numerical results of
Eq.~(\ref{d17}) and obtain rough estimates of the EM corrections.
The reasoning is that since the physical amplitudes are
independent of the scale $\mu$, there must be compensating
$\mu$-dependence between the chiral logarithms of Eq.~(\ref{d16})
and the counterterms.  Therefore the counterterms must be
at least of the same order-of-magnitude as the chiral logs
or even larger.  We have adopted the operational 
procedure of assuming the counterterm contribution 
$a_i^{\rm (ct)}$ vanishes at the scale 
$\mu = M_\rho$, and we assign an uncertainty given by $\pm |c_i|$.   
This leads to the numerical values 
\beqa
\delta (A_0^{\rm em}) &=& \left( 0.024 \pm 0.026 \right) \cdot
10^{-7}~M_{K^0} \ \ ,
\nonumber \\
\delta (A_2^{\rm em}) &=& \left( 0.015 \pm 0.022 \right) \cdot
10^{-7}~M_{K^0} \ \ ,
\nonumber \\
\delta (A_2^{+{\rm em}})  &=& \left( - 0.005 \pm 0.005 \right) \cdot
10^{-7}~M_{K^0} \ \ ,
\label{e1} \\
{\cal A}_{5/2} &=& \left( 0.012 \pm 0.016 \right) \cdot
10^{-7}~M_{K^0} \ \ ,
\nonumber
\eeqa
with $A_0 = (5.458 \pm 0.012) \cdot 10^{-7}~M_{K^0}$ and 
$A_2 = (0.2454 \pm 0.010) \cdot 10^{-7}~M_{K^0}$. 
Specifically, for the EM shift $\delta (A_2^{+{\rm em}}/A_2)$
calculated in Ref.~\cite{cdg}, we now have the extended result
\beq
{ \delta (A_2^{+{\rm em}}) \over A_2} \ = \
- \left( 2.0 \pm 2.2 \right) ~\% \ \ .
\label{e2}
\eeq
If one allows for the uncertainty in $g_{\rm emw}$
in addition to those in the counterterm values, we find
\beq
{ \delta (A_2^{+{\rm em}}) \over A_2} \ = \
- \left( 2.0^{+4.0}_{-2.2} \right) ~\% \ \ .
\label{e3}
\eeq
In the numerical findings of Eqs.~(\ref{e1})-(\ref{e3}),
the error bars are seen to be almost as large or larger
than the signal.  In our opinion, this is the best that one
can do within a strict chiral perturbation theory approach.

Our results illustrate several general features:
\begin{enumerate}
\item Since the
central values of the amplitudes have $\delta A_2^{em} \ne \delta
A_2^{+em}$, the electromagnetic loop corrections are seen to produce
$\Delta I = 5/2$ effects, although the uncertainties of the
counterterm values overwhelms the numerical result.
\item A phenomenological analysis~\cite{co} based on
$S$-wave pion-pion scattering lengths and forward dispersion
relations gives $\delta_0 - \delta_2 = (42 \pm 4)^o$.
Yet an isospin analysis of $K \to \pi\pi$ decays yields
$\delta_0 - \delta_2 = (56.7 \pm 3.9)^o$.  Presumably
this difference of nearly $15^o$ can be reconciled by
subtracting EM effects from the $K \to \pi\pi$ decays.
The main EM shift should be in $\delta_2$ as only this 
angle experiences a $\Delta I = 1/2$ enhancement.
Using Eq.~(\ref{d17a}) to calculate the angle $\gamma_2$
of Eq.~(\ref{a6}), we find
\beq
\gamma_2 = { {\cal A}_0^{(e^0p^2)} \over {\cal A}_2^{(e^0p^2)} }
\cdot {\beta \over 32 \pi} \left[ {\cal T}_{02}^{(e^2 p^0)}
- {2 \sqrt{2} \over 3 \beta^2} ~{\delta M_\pi^2 \over M_K^2} ~
{\cal T}_2^{(e^0 p^2)} \right] \simeq 4.5^o \ .
\label{e4}
\eeq
This evaluation, valid at order $e^2 p^0$, is seen to 
worsen the discrepancy between the two determinations.  
To reveal the explanation behind this puzzle requires 
more work.~\cite{cdg4}   
\item Finally, the most important implication of
these estimates is that the electromagnetic shifts in $A_2$ are not
large, being only a few percent. Naive estimates allow the
possibility that this shift could be much larger, perhaps even being a
major portion of $A_2$. Our previous work at the leading order in the
chiral expansion yielded a small effect. One motivation of the present
calculation was to see if the next order effects upset this
conclusion. Our estimates show that the natural size of the shift in
$A_2$ remains at the few percent level.
\end{enumerate}
This has been a complicated calculation with many different
lagrangians, describing different aspects of electromagnetic physics,
required to obtain the full effect. These include explicit photon
loops, mass shifts in the mesons propagating in loops and the
short-distance electroweak interaction. The chiral power counting was
crucial in sorting out which effects must be included for a consistent
calculation. The resulting structure is universal and model
independent. However, it is a prelude to more fully predictive
applications, as there remain unknown low energy constants which are
not predicted by chiral symmetry alone. Different models can be used
to estimate the renormalized constants which appear in the chiral
lagrangians, and these model predictions can then be readily
translated into the physical amplitudes through the use of our
calculation. In a following publication, we attempt to describe the
extent that this may be accomplished using dispersive techniques to
match long and short distance physics~\cite{cdg2}.

\vspace{1.cm}

The research described here was supported in part by the
National Science Foundation.  One of us (V.C.) acknowledges
support from M.U.R.S.T.  We thank John Belz for a useful 
communication.

\eject

\end{document}